\documentclass[prc,twocolumn,showpacs]{revtex4-2}

\usepackage{amsmath,amssymb,bm,bbm}
\usepackage{graphicx}
\usepackage{multirow}
\usepackage{dcolumn}
\usepackage{rotating}
\usepackage{xcolor}
\usepackage{tikz}
\usepackage{hyperref}
\usepackage{CJKutf8}
\usepackage{ulem}

\definecolor{green3}{rgb}{0.20,0.60,0.20}
\definecolor{Black}{rgb}{0.000000,0.000000,0.000000}
\definecolor{Blue}{rgb}{0.000000,0.000000,1.000000}
\definecolor{Cyan}{rgb}{0.000000,1.000000,1.000000}
\definecolor{Red}{rgb}{1.000000,0.000000,0.000000}
\definecolor{Green}{rgb}{0.000000,1.000000,0.000000}
\definecolor{Magenta}{rgb}{1.000000,0.000000,1.000000}
\definecolor{White}{rgb}{1.000000,1.000000,1.000000}
\definecolor{Yellow}{rgb}{1.000000,1.000000,0.000000}
\definecolor{Violet}{rgb}{0.552941,0.219608,0.788235}

\newcommand{\eq}[1]{\begin{equation} #1 \end{equation}}
\newcommand{\ket}[1]{ | #1 \rangle }

\newcommand{\elmx}[3]{\langle #1 | #2 | #3 \rangle}

\newcolumntype{.}{D{.}{.}{-1}}

\makeatletter
\def\p@subsection{}
\makeatother

\begin{document}

\begin{CJK*}{UTF8}{gbsn}


\title{$K$-isomeric states in the isotopic and isotonic chains of $^{178}$Hf}


\author{N. Minkov}
\email{nminkov@inrne.bas.bg}
\affiliation{Institute for Nuclear Research and Nuclear Energy, Bulgarian Academy of Sciences, Tzarigrad Road 72, BG-1784, Sofia, Bulgaria}

\author{L. Bonneau}
\email{bonneau@lp2ib.in2p3.fr}
\affiliation{LP2I Bordeaux, UMR 5797, Universit\'e de Bordeaux, CNRS, F-33170, Gradignan, France}

\author{P. Quentin}
\affiliation{LP2I Bordeaux, UMR 5797, Universit\'e de Bordeaux, CNRS, F-33170, Gradignan, France}

\author{J. Bartel}
\affiliation{IPHC, UMR 7178, Universit\'e de Strasbourg, CNRS, F-67000, Strasbourg, France}

\author{H. Molique}
\affiliation{IPHC, UMR 7178, Universit\'e de Strasbourg, CNRS, F-67000, Strasbourg, France}

\author{Meng-Hock Koh (辜 明 福)}
\affiliation{Department of Physics, Faculty of Science, Universiti
  Teknologi Malaysia, 81310 Johor Bahru, Johor, Malaysia}
\affiliation{UTM Centre for Industrial and Applied Mathematics, 81310
  Johor Bahru, Johor, Malaysia}

\date{\today}

%
%
%
%

\begin{abstract}
We study the evolution of $K^{\pi}=6^{+}$ and $8^{-}$ two-quasiparticle (q.p.) configurations in the isotopic and isotonic chains of even-even deformed nuclei around $^{178}$Hf and their ability to describe series of observed $K$-isomer excitations within the framework of a Skyrme Hartree--Fock--BCS (SHFBCS) approach using SIII interaction and seniority pairing strengths with self-consistent blocking. We apply the approach along the prescription in [Phys. Rev. C {\bf 105}, 044329 (2022)] used to describe $K$-isomers in the actinide and transfermium mass regions. The calculations allow us to identify the regions where proton or neutron configurations or their mixture may be responsible for the $K$-isomer formation. The obtained results provide a detailed test for the Skyrme SIII interaction used and outline the limits of applicability of the overall SHFBCS approach in the regions of well deformed nuclei. The study suggests that similar systematic analysis can be implemented in the heavier mass regions whenever enough data are available.
\end{abstract}

\maketitle

\end{CJK*}

%
%
%
%

\section{Introduction}
\label{introduction}

More than a century after their discovery nuclear isomers \cite{WD99,Herz08} remain one of the most exciting subjects in the nuclear structure study~\cite{WP20}. In particular, the $K$-isomers, for which a continuous stream of data comes from the nowadays advanced experimental facilities, provide new detailed information about the intrinsic shell-structure configurations which govern the appearance of nuclear metastable states offering at the same time a stringent test for the effective interactions used in the many-body theories of the nucleus. In the last decade the wealth of data firmly expands from the regions of moderate masses, such as the rare-earth and trans-lanthanide ($72 \leq Z \leq 82)$ nuclei~\cite{DRAC} (for recent works see \cite{Yokoyama17,Hartley,Petrache23}), to the region of very heavy and superheavy nuclei \cite{David15}--\cite{Chakma23} (for recent comprehensive review on trans-uranium nuclei see \cite{Hessberger23}). Up to date collections of data on nuclear isomers are available in \cite{NUBASE20} and \cite{atlas_isomer23}.

These experimental developments have been mirrored by a number of model descriptions within diverse theoretical approaches ranging from Nilsson-Strutinsky or Woods-Saxon mean fields \cite{WalXu16} (see references therein) to many-body self-consistent theories \cite{RS80}. In the former direction an advance in the $K$-isomer study was made through the so-called configuration-constrained energy surfaces in which the Nilsson quantum numbers of given single-particle (s.p.) configuration responsible for the isomer formation are traced through level crossings over the deformation surface (e.g. see \cite{Liu11,Liu13,Liu14} for applications to trans-actinides and superheavy nuclei). Further development in this direction was made by using the configuration-constrained rotation approach \cite{Fu13,Fu14} and the projected shell model (e.g. see \cite{Yang10,Chen13,Jiao15} for applications  in rare-earths and heavier nuclei) allowing for the study of rotation spectra built on $K$-isomeric states. More recent progress in the $K$-isomer study was made in the framework of the cranked Nilsson shell model with high-multipolarity deformations and particle-number conserving achieved through diagonalization of the Hamiltonian in an appropriately truncated cranked many-particle configuration space (see \cite{Zhang18,He18,He20} for applications to neutron-rich rare-earth nuclei and $^{254}$No). Regarding the Woods-Saxon potential, it was applied in a deformed shell model with pairing interaction to study the influence of octupole deformation on the two q.p. energy and magnetic moments in the $K$-isomeric states in the range from rare-earth to actinide and transfermium nuclei including $^{270}$Ds \cite{Minkov14}. An advanced use of the Woods-Saxon mean field was made in a macroscopic-microscopic model framework allowing for the prediction of a large amount of multi-q.p. configurations candidates for high-$K$ isomeric states in very heavy and superheavy nuclei \cite{Jach15,Jach18,Jach23}. A macroscopic-microscopic approach based on a two-center shell model has been also applied for the prediction of variety of $K$-isomeric states in superheavy nuclei \cite{Adamian10} as well as for the study of the Coriolis mixing effect on the isomer lifetimes in heavy nuclei \cite{Shneidman22}.

In the latter direction---many-body self-consistent theories---it has been shown that the energy-density-functional (EDF) theory, in its non-relativistic Skyrme and Gogny, as well as, relativistic (covariant density functional) realizations, is capable of reproducing q.p. spectra in heavy nuclei \cite{Dobaczewski15} (see references therein). A deformed Hartree-Fock (HF) approach with surface delta residual interaction and angular momentum projection was applied to describe known $K$-isomers---and predict possible unobserved ones---in the rare-earth mass region (Gd and Dy isotopes) \cite{Ghorui18}. In a recent work of present authors a Skyrme Hartree--Fock--BCS (SHFBCS) approach with SIII interaction and self-consistent blocking was applied to describe two q.p. energies and predict magnetic dipole moments in $K$-isomeric states of actinide and trans-fermium nuclei \cite{PRC22_hfbcs_isomers}. Also in a recent work, a Hartree-Fock-Bogoliubov (HFB) approach with density-dependent Gogny force used with both blocking and equal filling approximations was successfully applied to two- and four-q.p. $K$-isomeric states in $^{254}$No, $^{178}$Hf and several nuclei of the  Tungsten isotopic chain \cite{Robledo23}. An advanced application of the covariant density functional theory (CDFT) was made in works \cite{Prassa15,Karakat20} where relativistic Hartree-Bogoliubov (RHB) calculations with blocking and time-reversal symmetry breaking were performed for two-q.p. $K$-isomer configurations in trans-actinide nuclei around $N = 162$ \cite{Prassa15} and nuclei in the region from Er to Pb \cite{Karakat20}.

We remark that, while in the trans-actinide and superheavy nuclei the collection of data on $K$-isomers still needs to expand, in the rare-earth and trans-lanthanide nuclei the available data already encompass rather long continuous chains of isotopes and isotones \cite{DRAC}. The latter allow not only to follow the evolution of the particular proton or neutron configurations in the $K$-isomer structure, but also to examine the capability of the effective interactions to reproduce it as well as to explore the limits of applicability of the model approximations used. Such a study was made in the above mentioned Ref.~\cite{Karakat20} (in the CDFT framework), where the RHB calculations performed for the $K^{\pi}\!=\!6^{+}$ and $8^{-}$ isomeric energies in the Hafnium isotopic and $N=104,106$ isotonic chains provided a detailed test of the used density-dependent meson-exchange DD-ME2 and point-coupling DD-PC1 functionals. On the other hand, so far similar investigations within the non-relativistic EDF approach, especially using Skyrme effective interaction, have not been done in this mass region. Moreover, except for the recent SHFBCS study in the trans-actinide region \cite{PRC22_hfbcs_isomers}, to our knowledge no other specific application of the Skyrme EDF was made to investigate $K$-isomeric states. A number of works such as Ref.~\cite{Dobaczewski15} (and works quoted therein) describe a variety of q.p. excitations in nuclei of different mass regions, however, without particular focus on $K$-isomer properties (see also \cite{Bender03,Schunck10,Shi14}). In this aspect testing the Skyrme effective interaction on the long isotopic and isotonic series of $K$-isomeric states in the rare-earth nuclei would be important not only for assessing the force itself but also for achieving a new specific explanation of the observed and predicted systematic behavior of the s.p. excitations and nuclear shell structure in this mass region.

Motivated by the above, the purpose of the present work is to explore the series of $K$-isomeric states in the isotopic and isotonic chains to which $^{178}$Hf belongs with the use of the Skyrme EDF. Based on the success of our recent work on trans-actinide nuclei \cite{PRC22_hfbcs_isomers} we now assess for the first time the capability of the Skyrme effective interaction to reproduce the systematic behaviour of $K$-isomer energies and magnetic dipole moments in the long series of nuclei in the rare-earth and trans-lantanide region. We thoroughly examine the relevance of different proton and neutron two-q.p. configurations and their potential mixing to explain the formation mechanism and properties of the $K$-isomer excitations in dependence on the underlying nuclear shell structure.

We consider the series of $K^{\pi}\!=\!6^{+}$ and $8^{-}$ two-quasiparticle excitations in the isotopic chains of $^{168-184}$Hf and $^{170-186}$Hf, respectively, and the $K^{\pi}\!=\!8^{-}$ series in the $N\!=\!106$ isotonic chain from $^{170}$Gd to $^{182}$Os. As usual $K$ and $\pi$ denote respectively the projection of the total nuclear angular momentum on the symmetry axis in the body-fixed frame and the intrinsic parity. We apply a selfconsistent Skyrme--Hartree--Fock plus Bardeen-Cooper-Schrieffer (SHFBCS) approach with the SIII parametrization \cite{SIII} to calculate the $K^{\pi}\!=\!6^{+}$ and $8^{-}$ excitation energies using the same numerical algorithm as for the description of $K$-isomers in the regions of heavy actinide and transfermium nuclei~\cite{PRC22_hfbcs_isomers}. The configurations to be blocked are determined as follows:
\begin{itemize}
\item For the $K^{\pi}\!=\!6^{+}$ isomeric state we consider the
  two-neutron $(\frac 5 2^-[512], \frac 7 2^-[514])_n$ and the
  two-proton $(\frac 5 2^+[402], \frac 7 2^+[404])_p$ blocked
  configurations, which are the lowest two in the $^{178}$Hf nucleus,
  and follow them in even-$A$ Hf isotopes between $A\!=\!168$ and $A\!=\!186$.
\item As for the $K^{\pi}\!\!=\!\!8^{-}$ states, we take the two-neutron
  $(\frac 7 2^-[514], \frac 9 2^+[624])_n$ and the two-proton $(\frac
  72^+[404], \frac 9 2^-[514])_p$ blocked configurations, which are
  the lowest two in the $^{178}$Hf nucleus, and follow them in Hf
  isotopes between $A\!=\!168$ and $A\!=\!186$, and in $N\!=\!106$ even-$A$
  isotones between $A\!=\!170$ and $A\!=\!182$.
\end{itemize}
When relevant, we also consider energetically favorable alternate configurations (see Section~III).

As it will be shown below, the obtained series of theoretical energy levels and attendant magnetic moments compared to experimental data allow us to identify the configurations most probably contributing to the formation of the considered $K$-isomer excitations along the corresponding chains of nuclei. Also, the comparison with other theoretical approaches, such as the CDFT applied in the same mass region \cite{Karakat20}, allows us to assess on the same footing the applicability of the SIII Skyrme parametrization in the study of nuclear $K$-isomer excitations.

In Section II we briefly present the SHFBCS approach used together with some details on the pairing and basis parameters choice and the computational algorithm. Numerical results are presented and discussed in Section III, before presenting some concluding remarks in Section IV.

%
%
%
%

\section{Theoretical framework}

The theoretical approach used here includes a Skyrme HFBCS energy-density functional with cylindrical symmetry (axial deformation) in the intrinsic frame and self-consistent blocking of the single-particle orbitals entering the excited two-quasiparticle configuration \cite{PRC22_hfbcs_isomers}. We employ the SIII Skyrme parametrization \cite{SIII} within  a ``minimal'' scheme including the spin and current vector time-odd fields only, as explained in Ref.~\cite{Bonneau15}. The latter cause a time-reversal symmetry breaking at the one-body level in the excited states leading via the self-consistent blocking to a removal of the Kramers degeneracy in the single-particle energy spectrum.

The HF Hamiltonian is diagonalized and the single-particle spectrum is obtained as an expansion in the axially-deformed harmonic-oscillator basis \cite{DPPS} truncated at the $N_0+1 = 15$ major oscillator shell. The matrix elements are calculated through quadratures using 30 Gauss--Hermite mesh points in the $z$ and 15 Gauss--Laguerre mesh points in the perpendicular direction. The basis parameters $b$ and $q$ (defined in Ref.~\cite{DPPS}) are optimized for the considered nuclei. Because the resulting parameters vary only marginally over the considered chains of isotopes and isotones, with a very small effect on excitation energies of less than a couple of tens of keV (barely visible in the figures below) and on magnetic moments of a few thousands of the nuclear magneton $\mu_N$, we retain the same values $b=0.480$ and $q=1.185$ for all nuclei. Moreover this allows us to make comparisons between different nuclei with the same single-particle basis.

The pairing correlations are taken into account through the expectation values of a seniority residual interaction in the BCS states with a subsequent blocking at each iteration of the single-particle states entering the excited state configuration. The nucleon-number dependence of the corresponding matrix elements is parameterized as in Ref.~\cite{Bonneau15}. The BCS equations are solved for all single-particle states with a smearing factor $f(e_i) = \big[1+\exp\big((e_i-X-\lambda_{\tau})/\mu\big)\big]^{-1}$ where $e_i$ is the energy of the single-particle state $\ket i$, $X \!=\! 6\;$MeV is a truncation parameter, $\lambda_{\tau}$ is the chemical potential for $\tau =n,p$ (neutrons and protons) and $\mu \!=\! 0.2$~MeV is a diffuseness parameter. The neutron and proton pairing-strength constants are taken for all considered nuclei as $G_n=16$~MeV and $G_p=15$~MeV based on earlier proposed concept of overall adjustment with respect to experimental estimations for the moment of inertia \cite{consistency} (see also Section IV in~Ref.~\cite{PRC22_hfbcs_isomers}).

The theoretical two-quasiparticle excitation energy is determined as
\begin{eqnarray}
E^{*}_{\rm th}(K^{\pi})= E_{\mbox{\scriptsize
    tot}}^{2qp}(K^{\pi})-E_{\mbox{\scriptsize tot}}^{\mbox{\scriptsize
    GS}} \ ,
\label{Eiso}
\end{eqnarray}
where $E_{\mbox{\scriptsize tot}}^{2qp}(K^{\pi})$ is the total HFBCS energy of the nucleus, obtained in the solution with blocked isomer configuration orbitals and $E_{\mbox{\scriptsize tot}}^{\mbox{\scriptsize GS}}$ is the total energy in the ground-state solution. We emphasize that here the $K$-isomer excitation energy is not simply the sum of the quasiparticle energies of two neutrons/protons that would be calculated from the GS solution. As a difference between the total energies obtained in the solutions for the blocked 2q.p. configuration and the ground state, $E^{*}_{\rm th}(K^{\pi})$ incorporates the relevant nuclear bulk properties determined in the self-consistent calculations. Also, we note that although the HFBCS is not a configuration-mixing approach it takes into account the quasiparticle interaction through the one-body reduction of the two-nucleon potential. The same is valid for the CDFT approach used in \cite{Prassa15,Karakat20}. As mentioned in Sec.~\ref{introduction} the relevance of the Skyrme EDF for the description of q.p. excited bandheads was pointed out in a number of earlier works (e.g. see Refs.~\cite{Dobaczewski15,Bender03,Schunck10,Shi14}), while the particular use of the Skyrme SIII parametrization for description of $K$-isomer excitations in trans-actinide nuclei was demonstrated in our recent work \cite{PRC22_hfbcs_isomers}. Moreover, the latter was proved to be capable of describing $K$-isomeric states in the presence of octupole deformation \cite{BJP2019}.

The magnetic dipole moments in the two-quasiparticle configurations are calculated by taking into account core-polarization effects as explained in Ref.~\cite{Bonneau15}. In the same reference, the particular choice made to estimate the collective gyromagnetic ratio (as calculated in Eq.\ (13) of Ref.~\cite{Bonneau15}) is detailed:
(i) the BCS pairs are built from s.p.\ canonical basis states which are almost time-conjugated,
(ii) the blocked pairs are removed from the summations,
(iii) the same quasi-particle energy is used for both members of a BCS pair.

%
%
%
%
%
%

\begin{figure}[ht]
\centering
\includegraphics[width=0.480\textwidth]{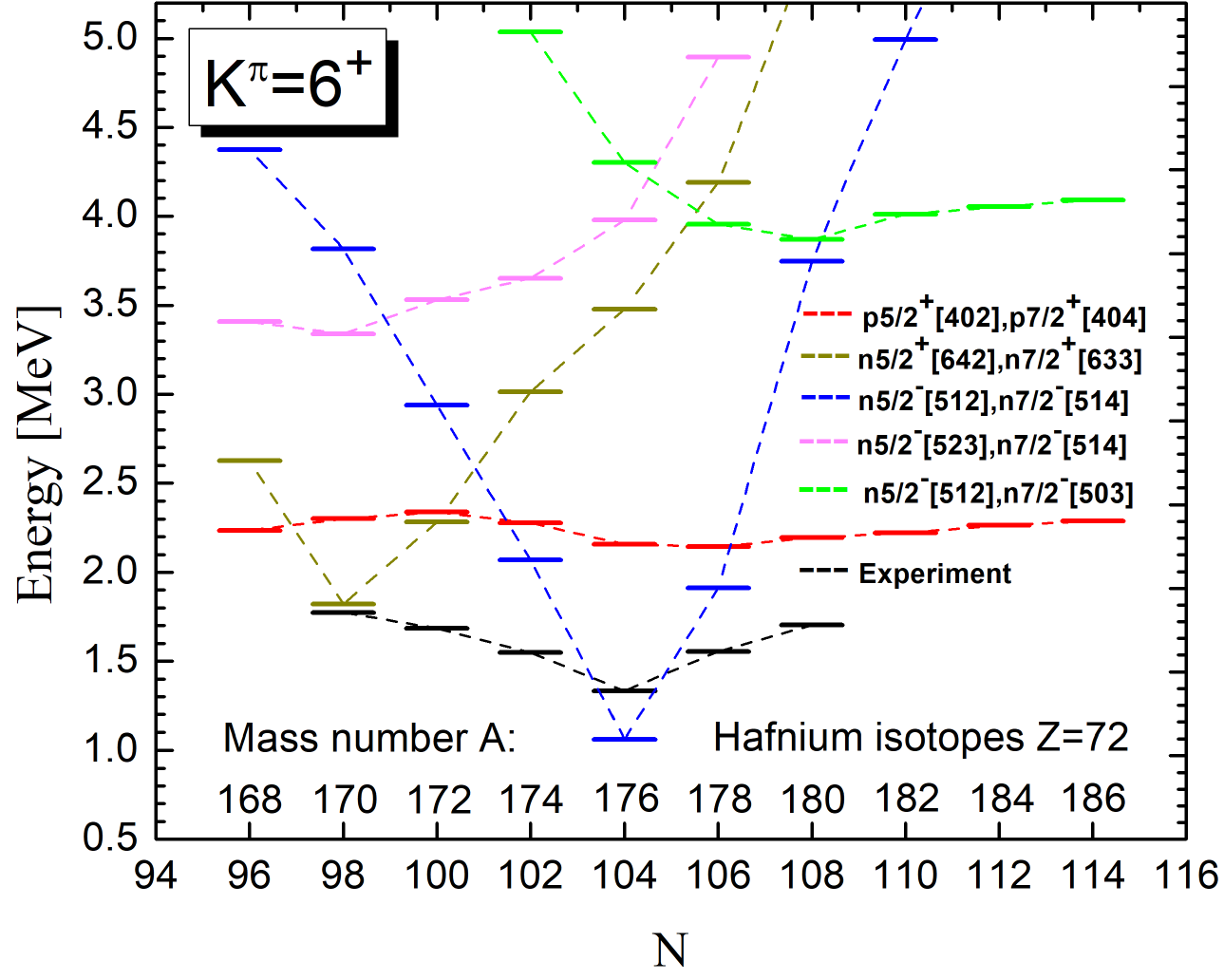}
\caption{Theoretical $K^{\pi}=6^{+}$ excitation energies of low-lying two-neutron and two-proton blocked configurations obtained in even Hf isotopes from $A=168$ to $A=186$ compared with experimental data (in black) \cite{DRAC}.}
\label{hf168-184_6plus}
\end{figure}

\section{Numerical results and discussion}

We performed numerical SHFBCS calculations for the $K^{\pi}=6^{+}$ and $8^{-}$ lowest-energy neutron and proton two-quasiparticle configurations in the isotopic and isotonic chains of well-deformed nuclei around $^{178}$Hf, as explained above. The results are given in Tables~\ref{tab:hafnium_eismagmom} and \ref{tab:n106isot_eismagmom}, where the deformation characteristics of the considered isotopes/isotones is expressed by the experimental $R_{42}=E(4_{1}^{+})/E(2_{1}^{+})$ ratio of the energies of the first two members of the ground-state rotational band. For a good rotator the $R_{42}$ value should be larger or of the order of 3.2. For each nucleus we compare the obtained neutron and proton two-quasiparticle excitation energy $E^{*\tau}_{\rm th}(K^{\pi})$ ($\tau=n,p$) with the experimental value for the corresponding $K$-isomeric state \cite{DRAC,NUBASE20}. In a few cases, where no data are available, we still give the theoretical predictions for better understanding of the evolution of isomeric energies along the considered isotopic and isotonic chains. In the tables we also give theoretical magnetic dipole moments calculated for each two-quasiparticle configuration and compare them with experimental data where available.

It should be noted that the considered Hf isotopes are part of the region of collective rotation with an experimental $R_{42}$ ratio varying between $3.11-3.19$ in $^{168-170}$Hf and a value of 3.31 in $^{180}$Hf \cite{NNDC} (see Table~\ref{tab:hafnium_eismagmom}). The corresponding quadrupole deformation parameter $\beta$ takes values $\beta\approx 0.27-0.28$ in $^{168}$Hf and $^{180}$Hf and $\beta\approx 0.30-0.31$ in $^{170-176}$Hf \cite{Prit16}. In the considered $N=106$ even isotones, the good-rotor character is experimentally confirmed only from $^{176}$Yb to $^{180}$W ($R_{42} \sim 3.3$), whereas the $R_{42}$ value is somewhat lower in $^{182}$Os (about 3.15) and experimental data for $^{170}$Gd up to $^{174}$Er are unavailable in Ref.~\cite{NNDC}.

\begin{table*}[ht]
\renewcommand{\arraystretch}{1.3}
\begin{center}
  \caption{Theoretical (HFBCS) excitation energies
    $E^{*n}_{\rm th}$ and $E^{*p}_{\mbox{\scriptsize
      th}}$ (in MeV) and magnetic dipole moments
    $\mu_{\rm th}^{n}$ and $\mu_{\mbox{\scriptsize
        th}}^{p}$ (in nuclear magneton unit) for the lowest-energy neutron
    and lowest-energy proton $K^{\pi}=6^{+}$ and $K^{\pi}=8^-$
    configurations in even Hf ($Z=72$) isotopes from $A=168$ to
    $A=186$, compared with the experimental $K^{\pi}=6^{+}$
    \cite{DRAC} and $K^{\pi}=8^{-}$ \cite{NUBASE20} isomeric energies
    (except for the $8^-$ isomeric energy in $^{170}$Hf taken from
    Ref.~\cite{atlas_isomer23}) Magnetic dipole moments are taken
    from Refs.~\cite{Stone19,Stone20} (with the published
    references). Experimental $R_{42}$-ratios \cite{NNDC} are
    also given (third column). Blank entries correspond to
    missing data. \label{tab:hafnium_eismagmom}}
  \begin{tabular}{*{9}crrr}
\hline \hline
$A$ & $N$ &$R_{42}$& $K^{\pi}$ & Neutron configuration
& $E^{*n}_{\rm th}$ & Proton configuration
& $E^{*p}_{\rm th}$ & $E^{*}_{\mbox{\scriptsize exp}}$ &
\multicolumn{1}{c}{$\mu_{\rm th}^{n}$} &
\multicolumn{1}{c}{$\mu_{\rm th}^{p}$} &
\multicolumn{1}{c}{$\mu_{\mbox{\scriptsize exp}}$} \\

\hline\hline
168 & 96 & 3.110 & $6^{+}$ & $\big(5/2^+[642],7/2^+[633]\big)$
& 2.626 & $\big(5/2^+[402],7/2^+[404]\big)$
& 2.236 &  & $-$1.249 & 5.740 &  \\
& & & $8^{-}$ & $\big(7/2^-[514],9/2^+[624]\big)$
& 5.277 & $\big(7/2^+[404],9/2^-[514]\big)$ & 2.348 &  &
0.251 & 7.399 & \\
\hline

170 & 98 & 3.194 & $6^{+}$ & $\big(5/2^+[642],7/2^+[633]\big)$
& 1.821 & $\big(5/2^+[402],7/2^+[404]\big)$
& 2.301 & 1.773 & $-$1.357 & 5.652 & \\
& & & $8^{-}$ & $\big(7/2^-[514],9/2^+[624]\big)$
& 4.663 & $\big(7/2^+[404],9/2^-[514]\big)$
& 2.265 & 2.183 & 0.148 & 7.348 & \\
\hline

172 & 100 & 3.248 & $6^{+}$ & $\big(5/2^+[642],7/2^+[633]\big)$
& 2.283 & $\big(5/2^+[402],7/2^+[404]\big)$
& 2.339 &1.685& $-$1.284 & 5.670 & +5.6(6)\cite{NPA349} \\
& & & $8^{-}$ & $\big(7/2^-[514],9/2^+[624]\big)$
& 3.722 & $\big(7/2^+[404],9/2^-[514]\big)$
& 2.129 & 2.006 & 0.062 & 7.338 & +7.93(6)\cite{NPA349} \\
\hline

174 & 102 & 3.268 & $6^{+}$ & $\big(5/2^-[512],7/2^-[514]\big)$
& 2.069 & $\big(5/2^+[402],7/2^+[404]\big)$
& 2.279 & 1.549 & 0.073 & 5.676 & +5.40(5)\cite{NPA349}\\
& & & $8^{-}$ & $\big(7/2^-[514],9/2^+[624]\big)$
& 2.774 & $\big(7/2^+[404],9/2^-[514]\big)$
& 1.910 & 1.798 & 0.150 & 7.335 &  \\
\hline

176 & 104 & 3.284 & $6^{+}$ & $\big(5/2^-[512],7/2^-[514]\big)$
& 1.061 & $\big(5/2^+[402],7/2^+[404]\big)$
& 2.159 & 1.333 & $-$0.029 & 5.697 &  \\
& & & $8^{-}$ & $\big(7/2^-[514],9/2^+[624]\big)$
& 1.737 & $\big(7/2^+[404],9/2^-[514]\big)$
& 1.598 & 1.559 & 0.267 & 7.342 & 7.93 \\
\hline

178 & 106 & 3.291 & $6^{+}$ & $\big(5/2^-[512],7/2^-[514]\big)$
& 1.913 & $\big(5/2^+[402],7/2^+[404]\big)$
& 2.143 & 1.554 & 0.106 & 5.703 & +5.81(5)\cite{NPA349} \\
& & & $8^{-}$ & $\big(7/2^-[514],9/2^+[624]\big)$
& 1.136 & $\big(7/2^+[404],9/2^-[514]\big)$
& 1.349 & 1.147 & 0.312 & 7.349 & +3.09(1)\cite{PLB645} \\
\hline

180 & 108 & 3.307 & $6^{+}$ & $\big(5/2^-[512],7/2^-[514]\big)$
& 3.748 & $\big(5/2^+[402],7/2^+[404]\big)$
& 2.194 & 1.703 & 0.168 & 5.695 &  \\
& & & $8^{-}$ & $\big(7/2^-[503],9/2^+[624]\big)$
& 3.009 & $\big(7/2^+[404],9/2^-[514]\big)$
& 1.179 & 1.142 & $-$1.922 & 7.342 & +8.7(10)\cite{PRL27} \\
\hline

182 & 110 & 3.295 & $6^{+}$ & $\big(5/2^-[512],7/2^-[503]\big)$
& 4.010 & $\big(5/2^+[402],7/2^+[404]\big)$
& 2.223 &   & $-$2.050 & 5.755 &  \\
& & & $8^{-}$ & $\big(7/2^-[503],9/2^+[624]\big)$
& 3.204 & $\big(7/2^+[404],9/2^-[514]\big)$
& 1.081 & 1.173 & $-$1.879 & 7.378 &  \\
\hline

184 & 112 & 3.264 & $6^{+}$ & $\big(5/2^-[512],7/2^-[503]\big)$
& 4.052 & $\big(5/2^+[402],7/2^+[404]\big)$
& 2.264 &  & $-$2.042 & 5.822 &  \\
& & & $8^{-}$ & $\big(7/2^-[503],9/2^+[624]\big)$
& 3.228 & $\big(7/2^+[404],9/2^-[514]\big)$
& 1.024 & 1.272 & $-$1.929 & 7.414 &  \\
\hline

186 & 114 &  & $6^{+}$ & $\big(5/2^-[512],7/2^-[503]\big)$
& 4.091 &  $\big(5/2^+[402],7/2^+[404]\big)$
& 2.290 &  & $-$2.021 & 5.895 &   \\
& & & $8^{-}$ & $\big(7/2^-[503],9/2^+[624]\big)$
& 3.209 & $\big(7/2^+[404],9/2^-[514]\big)$
& 1.057 &   & $-$2.032 & 7.455 & \\
\hline \hline
\end{tabular}
\end{center}
\end{table*}

\begin{table}[ht]
\renewcommand{\arraystretch}{1.2}
\begin{center}
\caption{Theoretical (HFBCS) excitation energies
    $E^{*n}_{\rm th}$ and $E^{*p}_{\mbox{\scriptsize
      th}}$ (in MeV) and magnetic dipole moments
    $\mu_{\rm th}^{n}$ and $\mu_{\mbox{\scriptsize th}}^{p}$
    (in nuclear magneton unit) for the lowest-energy $K^{\pi}=8^-$
  neutron $\big(\frac72^-[514],\frac 92^+[624]\big)$, and proton
  $\big(\frac72^+[404],\frac 92^-[514]\big)$ configuration, in even $N=106$ isotones
  from $A=170$ to $A=182$, compared with the experimental
  $K^{\pi}=8^{-}$ isomeric energies~\cite{DRAC}
  and magnetic dipole moments given by
  Refs.~\cite{Stone19,Stone20} (with the published
  references). Experimental $R_{42}$-ratios \cite{NNDC} are also
  given. Blank entries correspond to missing data.
  \label{tab:n106isot_eismagmom}}
\begin{tabular}{*{5}crrr}
\hline \hline
Nucleus & $R_{42}$ & $E^{*n}_{\rm th}$
& $E^{*p}_{\rm th}$ & $E^{*}_{\mbox{\scriptsize exp}}$ &
\multicolumn{1}{c}{$\mu_{\rm th}^{n}$} &
\multicolumn{1}{c}{$\mu_{\rm th}^{p}$} &
\multicolumn{1}{c}{$\mu_{\mbox{\scriptsize exp}}$} \\

\hline\hline
$^{170}_{\;\,64}$Gd &  & 1.118 & 6.746 &    & 0.331 & 7.491 \\
$^{172}_{\;\,66}$Dy &  & 1.060 & 5.629 & 1.278 & 0.356 & 7.520 \\
$^{174}_{\;\,68}$Er &  & 1.018 & 4.328 & 1.112 & 0.334 & 7.571 \\
$^{176}_{\;\,70}$Yb & 3.31 & 1.028 & 3.012 & 1.050 & 0.334 & 7.422 &
$-0.151(15)$\cite{PLB645}\\
$^{178}_{\;\,72}$Hf & 3.291& 1.136 & 1.349 & 1.147 & 0.312 & 7.349 &
+3.09(1)\cite{PLB645}\\
$^{180}_{\;\,74}$W  & 3.26 & 1.276 & 2.465 & 1.529 & 0.256 & 7.354 \\
$^{182}_{\;\,76}$Os & 3.154 & 1.365 & 3.314 & 1.831 & 0.253 & 7.391 \\
\hline \hline
\end{tabular}
\end{center}
\end{table}

\subsection{Isomeric energies}

\begin{figure*}[ht]
\centering
\includegraphics[height=0.55\textheight]{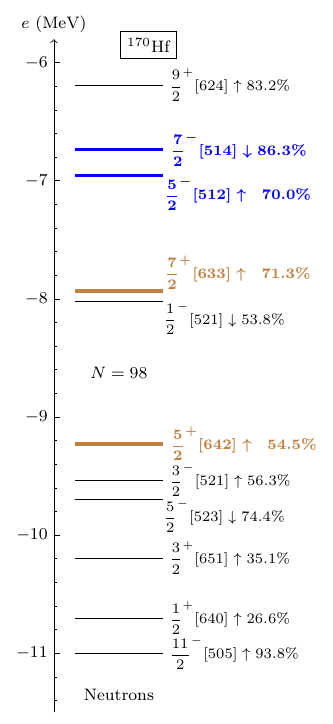}
\includegraphics[height=0.55\textheight]{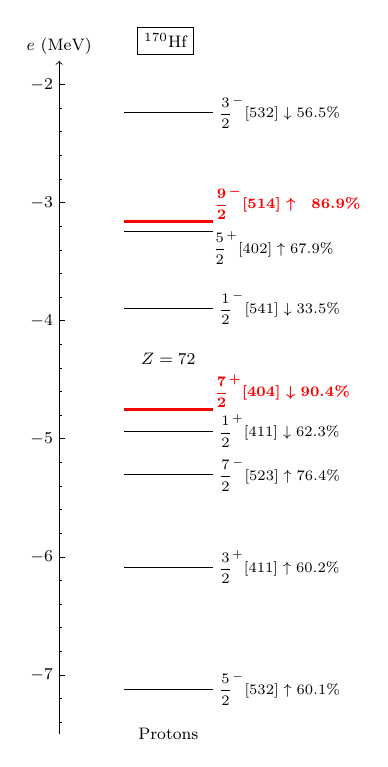}
\caption{Neutron and proton single-particle spectra in the ground-state solution of $^{170}$Hf. Upward and downward arrows denote spin projections on the symmetry axis equal $+1/2$ and $-1/2$, respectively, and the percentages given to their right is the weight of the corresponding Nilsson quantum numbers in the single-particle wave function. The boldface neutron levels in blue correspond to the lowest-energy $(\frac 52^-[512],\frac72^-[514])$ neutron configuration of the $K^{\pi}=6^+$ state in $^{178}$Hf and neighboring even Hf isotopes, whereas the brown levels correspond to the lowest-energy neutron configuration in the $^{170,172}$Hf isotopes. The boldface red proton levels identify the lowest-energy $K^{\pi}=8^-$ proton configuration along the considered chain of even Hf isotopes.}
\label{sp_spectra_Hf170}
\end{figure*}

\begin{figure}[ht]
\hspace*{-0.2cm}
\includegraphics[width=0.45\textwidth]{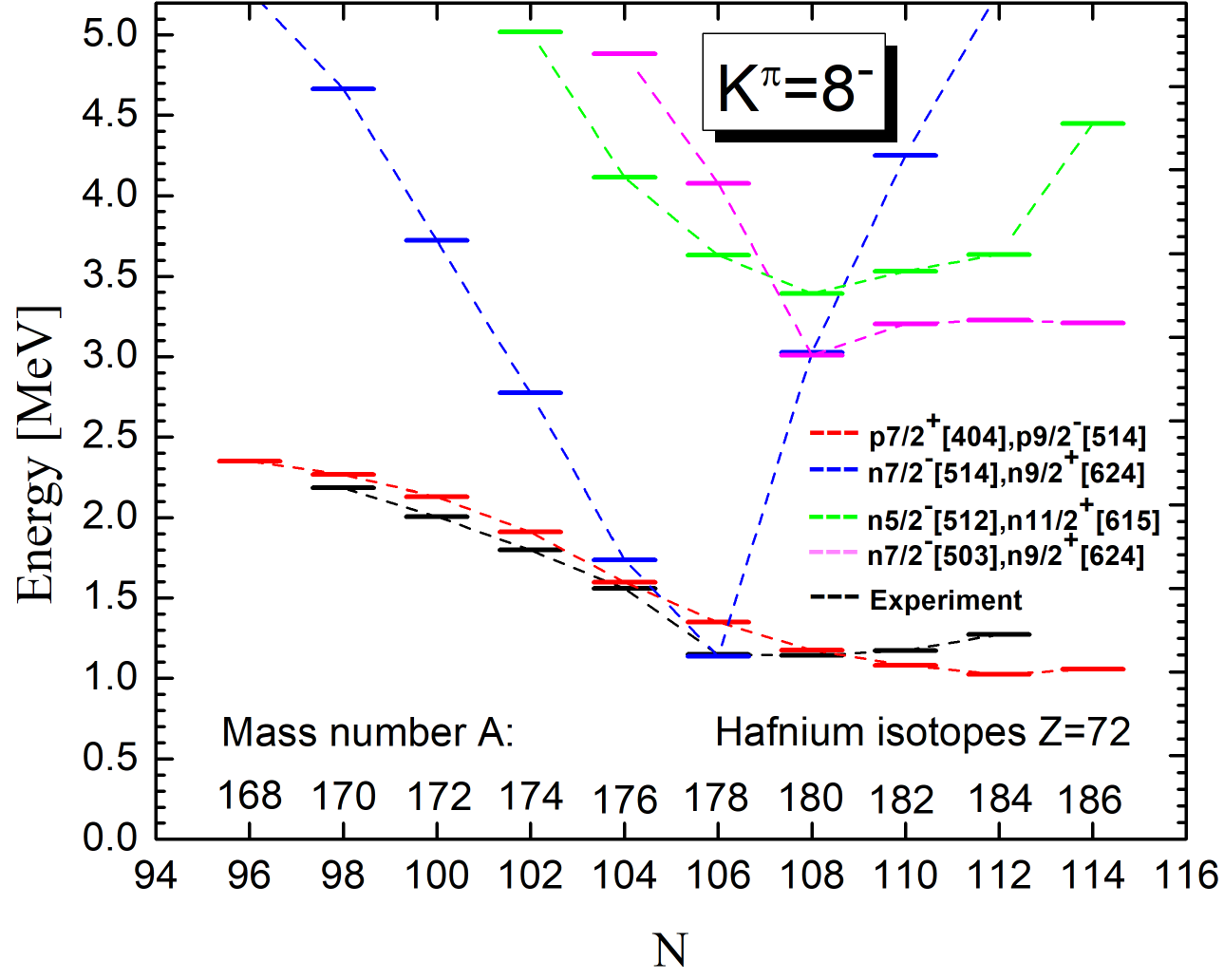}
\caption{Theoretical $K^{\pi}=8^{-}$ excitation energies of low-lying two-neutron and two-proton blocked configurations obtained in even Hf isotopes from $A=168$ to $A=186$ compared with experimental data (in black) \cite{NUBASE20}.}
\label{hf170-186_8minus}
\end{figure}

\begin{figure}[ht]
\centering
\includegraphics[width=0.45\textwidth]{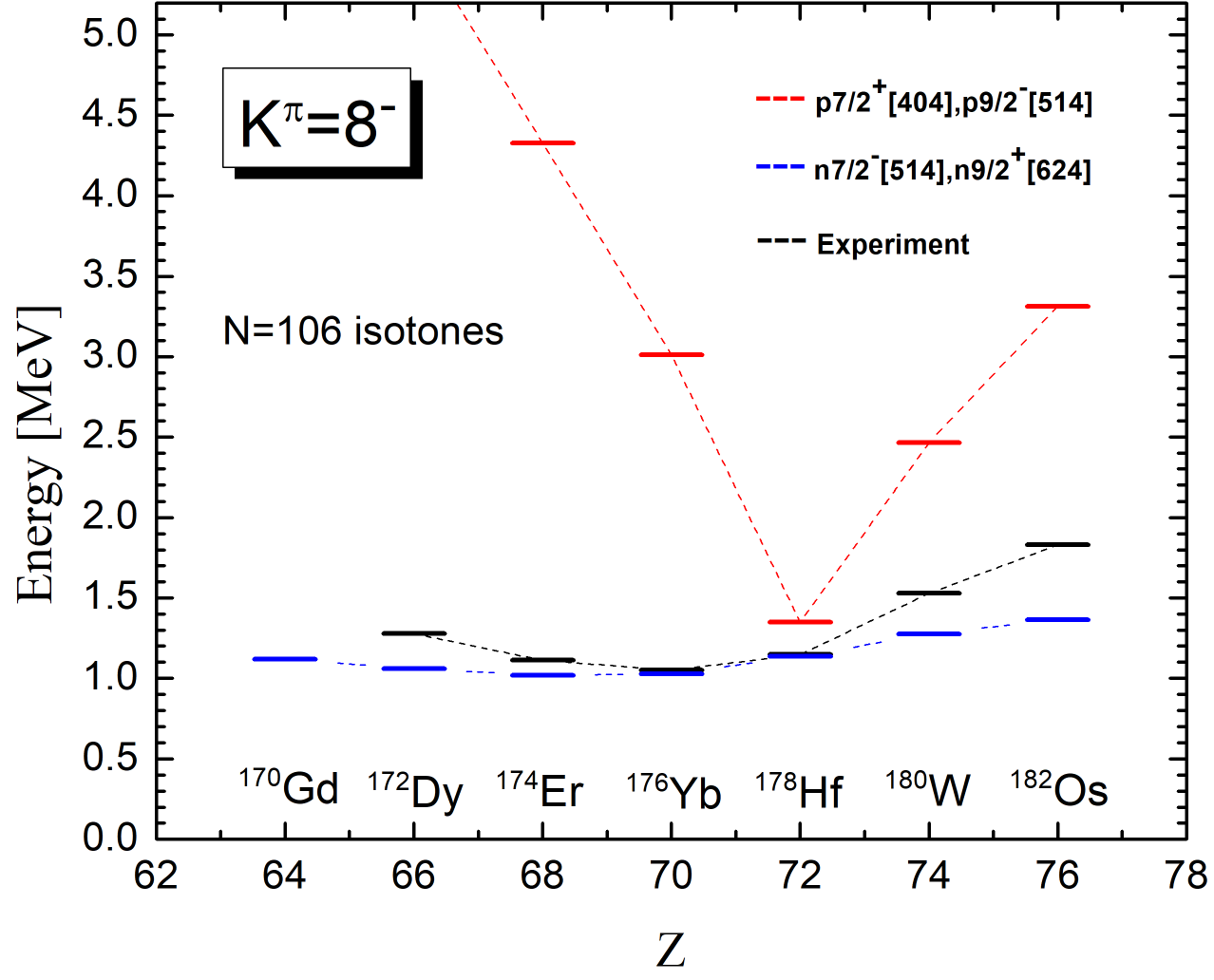}
\caption{Theoretical $K^{\pi}=8^{-}$ two-quasiparticle excitation energies obtained from neutron (blue bars) and proton (red bars) two-quasiparticle configurations for the $N=106$  isotonic chain between $^{170}$Gd and $^{182}$Os compared with experimental data (in black) \cite{NUBASE20}.}
\label{n=106isotones_8minus}
\end{figure}

The energy levels obtained in our calculations for $K^{\pi}=6^{+}$ excitations in the $^{168-186}$Hf isotopic chain are plotted in Fig.~\ref{hf168-184_6plus} as function of the neutron number. One finds that the energies corresponding to the two-proton $(\frac 5 2^+[402],\frac72^+[404])_p$ blocked configuration (red bars) follow the overall behavior of experimental data, overestimating them by between 0.5 and 0.8 MeV (see also Table~\ref{tab:hafnium_eismagmom}). On the other hand the energy levels obtained with the two-neutron $(\frac52^-[512],\frac72^-[514])$ blocked configuration (blue bars) approach the experimental levels more closely in $^{174-178}$Hf isotopes, but yield much too high excitation energies for lighter and heavier Hf isotopes. In $^{176}$Hf the theoretical value of $E^{*n}_{\rm th}(6^{+})=1.061$ MeV underestimates the experimental value of 1.333 MeV by 272 keV while in $^{174}$Hf and $^{178}$Hf the calculations overestimate the experimental value by 520 keV and 359 keV, respectively (see Table~\ref{tab:hafnium_eismagmom}). One may thus consider that in $^{176}$Hf the structure of the $6^{+}$ isomer is dominated by the two-neutron $(\frac 5 2^-[512], \frac 7 2^-[514])$ configuration while for $^{174}$Hf and $^{178}$Hf some admixture with the two-proton $(\frac 5 2^+[402], \frac 7 2^+[404])$ configuration is likely to occur. The latter has actually been found experimentally long time ago from two-nucleons transfer reactions in $^{178}$Hf with a dominance (69\%) of the ($\frac 52^{+}\!, \frac 72^{+}$) proton configuration with the Nilsson quantum numbers $[402]$ and $[404]$ (see Table I of Ref.\ \cite{KHOO}). Further away from these three Hf isotopes, the two-neutron $(\frac 5 2^-[512], \frac 7 2^-[514])$ configuration is definitely off, while the two-proton configuration still follows the experiment from above.

One notices that in the above considered $K^{\pi}\!=\!6^{+}$ excitations the leading components of the calculated single-particle wave functions in the axially-deformed harmonic-oscillator expansion, namely the $(\frac 52^-[512],\frac 72^-[514])$ for the two-neutron and the $(\frac 52^+[402],\frac 72^+[404])$ for the two-proton configuration, agree with the Nilsson quantum numbers proposed earlier in the literature (see, e.g., \cite{NNDC} for references). However, we find an alternative configuration to account for the $K^{\pi}=6^{+}$ isomer in the $^{170,172}$Hf isotopes, namely the neutron $(\frac 52^+[642],\frac 72^+[633])$ two-quasiparticle configuration, as can be seen in Fig.~\ref{sp_spectra_Hf170} in the single-particle spectra of $^{170}$Hf. A remarkable agreement with experiment is in particular obtained for this configuration in $^{170}$Hf.

Here it is interesting to make a comparison with the analogous result obtained in Ref.~\cite{Karakat20} through RHB calculations with DD-ME2 and DD-PC1 functionals. First, comparing our Fig.~\ref{hf168-184_6plus} with Fig.~6 in \cite{Karakat20} (RHB with DD-ME2) we notice a similarity in the shapes of the energy curves for the two-neutron $(\frac 52^+[642],\frac 72^+[633])$, $(\frac 52^-[512],\frac 72^-[514])$ and two-proton $(\frac 52^+[402],\frac 72^+[404])$ configurations as functions of the neutron number $N$. However, in the RHB calculation the ``neutron'' curves show much shallower minimum with $N$ while the ``proton'' curve appears higher in energy compared to our SHFBCS calculation. Then, comparing the energies of the lowest configurations obtained for the different isotopes in each calculation, our Table~\ref{tab:hafnium_eismagmom} and Table~II in \cite{Karakat20} (also compare our Fig.~\ref{hf168-184_6plus} with Fig.~7 in \cite{Karakat20}), we see that for $^{170}$Hf and $^{176}$Hf the $6^{+}$ energies obtained in the two approaches are rather similar. For $^{172}$Hf and $^{174}$Hf the RHB calculations provide better agreement with the experiment while for $^{178}$Hf and $^{180}$Hf better agreement is obtained in our SHFBCS calculation. The overall quality of the two model descriptions of the $K^{\pi}\!=\!6^{+}$ excitations in Hafnium isotopes looks similar, which points to the similar spectroscopic characteristics of the interactions used in the two approaches. Further detailed comparison of the Skyrme SIII HFBCS approach and the RHB approach with DD-ME2 and DD-PC1 on the basis of $K$-isomer description might be valuable but is beyond the scope of this work.

Next, we investigate the $K^{\pi}=8^{-}$ state with the two-neutron $(\frac 7 2^-[514], \frac 9 2^+[624])$ and two-proton $(\frac 7 2^+[404], \frac 9 2^-[514])$ blocked configurations. The obtained energy levels are plotted in Fig.~\ref{hf170-186_8minus} as functions of $N$. Similarly to the $6^{+}$ isomer, we observe that the energy of the two-proton configuration closely follows the experimental energies, with an especially good agreement in $^{176}$Hf and $^{180}$Hf, with an overestimation of the experimental isomeric energies of only about 40 keV, as can be seen in Table~\ref{tab:hafnium_eismagmom}. At the same time the two-neutron configuration compares favorably with experiment only in the $^{176}$Hf and $^{178}$Hf isotopes, with an overestimation of about 180 keV in $^{176}$Hf and an underestimation of only about 10 keV in $^{178}$Hf. In the other isotopes the theoretical two-neutron blocked configuration energy is obtained high above the experimental energy and anyway at excitation energies where a description of excited states as being of a pure 2qp nature is most doubtful.
Moreover one may note that for $A\geqslant 180$, the two-neutron $(\frac 7 2^-[503], \frac{9}2^+[624])$ configuration turns out to be  energetically more favorable than the $(\frac 7 2^-[514],\frac 9 2^+[624])$ configuration.

For $N = 106$ the Fermi level lies in between the two $\frac 72^{-} $ and $\frac 92^{+}$ neutron s.p.\ levels relevant to form a $8^{-}$ 2qp configuration. The corresponding excitation energy lies slightly below the proton ($\frac 72^{+}\!, \frac 92^{-}$) configuration describing adequately the experimental $8^{-}$ isomeric states below $N= 104$ and above $N =106$. As already pointed out long ago (see \cite{KHOO} and references quoted therein) the neutron-proton residual interaction yields a mixing of these two $8^{-}$ states in $^{178}$Hf resulting in two states, the isomeric one at 1147 keV and another 338 keV above. As pointed out in Ref. \cite{KHOO} (see their Table I) the isomeric $8^{-}$ state is considered from two nucleons transfer data to be mostly of a neutron nature (64\%). The fact that our neutron unperturbed state energy is lower than its proton counterpart points rightly in this direction.

Moving away from $ N = 106 $, apart from $ N = 104$, the location of the Fermi level in the neutron s.p.\ spectra raises considerably the excitation energy of this 2qp neutron $8^{-}$ configuration disqualifying it to describe the $8^{-}$ isomer. In $^{176}$Hf the quasi degeneracy of the neutron $\frac 72^+$ and $\frac 12^-$ s.p.\ levels does not lower much the Fermi level below its $N= 106 $ position allowing for a mixing of the two neutron and proton configurations in the $8^{-}$ isomeric state where its proton component should probably be dominant in view of its lower unperturbed energy as compared to the neutron one.

The best reproduction of the $8^{-}$ isomeric energy is thus achieved in $^{176,178,180}$Hf which also represent the best rotators in the isotopic chain with $R_{42}=3.28-3.31$ (see Table~\ref{tab:hafnium_eismagmom}). For all above considered $K^{\pi}=8^{-}$ excitations the two-neutron configuration was obtained with the two-neutron $(\frac72^-[514],\frac92^+[624])$ configuration, while for the two-proton configuration we have $(\frac 72^+[404], \frac 92^-[514])$, again corroborating earlier proposed Nilsson quantum numbers in the structure of these configurations (see \cite{NNDC} and references therein). The above results outline an overall region of the most relevant applicability in the Hf isotopic chain of the SHFBCS approximation with SIII interaction.
Here comparing again with Ref.~\cite{Karakat20} we notice a rather clear similarity in the shapes of the curves obtained for the energy of the two-proton configuration $(\frac 72^+[404], \frac 92^-[514])$ as a function of $N$ in our calculation (Fig.~\ref{hf170-186_8minus}) and in the RHB calculations with DD-ME2 and DD-PC1 functionals (Fig.~11 in \cite{Karakat20}). Also, we notice the overall similarity in the quality of the two model descriptions (compare our Table~\ref{tab:hafnium_eismagmom} and Table~IV in \cite{Karakat20}).

In Fig.~\ref{n=106isotones_8minus} the excitation energies corresponding to the $K^{\pi}=8^{-}$ isomer in the $N=106$ isotones are given as function of the proton number $Z$. One notices that the theoretical $E^{*n}_{\rm th}(8^{-})$ energy of the two-neutron $(\frac 7 2^-, \frac 9 2^+)_n$ configuration is closer to the experimental isomer energies as compared to the two-proton configuration $(\frac 7 2^+, \frac 9 2^-)_p$ which approaches the experiment only in $^{178}$Hf (overestimating it by 202 keV), while jumping to irrelevantly high values otherwise (see Table~\ref{tab:n106isot_eismagmom} and discussion above). The two-neutron configuration, on the contrary, provides a rather good agreement with the data for $^{176}$Yb and for the above mentioned case of $^{178}$Hf with an underestimation of respectively 22 and 11 keV. Aside of these two nuclei the disagreement between theory and experiment varies between a 90 and 470 keV underestimation along the isotonic chain from $^{172}$Dy to $^{182}$Os (see Table~\ref{tab:n106isot_eismagmom}). We remark that from $^{176}$Yb to $^{182}$Os the $R_{42}$ ratio takes values between 3.31 and 3.15, which indicates their presence in the rotation part of the rare earth region. No experimental information on $R_{42}$ is available in \cite{NNDC} for $^{170}$Gd, $^{172}$Dy and $^{174}$Er, but one may guess that it should be close to the rotation values. In all considered $N=106$ isotones we obtain the same Nilsson quantum numbers structure of the $K^{\pi}=8^{-}$ two-neutron and two-proton configurations as in the case of the Hf isotopic chain discussed above. We may thus conclude that the present calculations provide reasonably motivated model predictions for the $K^{\pi}=8^{-}$ isomer from $^{170}$Gd to $^{180}$W which identifies the corresponding region of applicability of the present SHFBCS approach with the SIII parametrization. Here the comparison with the RHB calculations of Ref.~\cite{Karakat20} shows the better quality of the present SIII HFBCS description (compare our Table~\ref{tab:n106isot_eismagmom} and Table~V in Ref.~\cite{Karakat20}). Finding the source of discrepancy between the two approaches in the case would be an interesting subject of a separate study.

\subsection{Magnetic dipole moment}

Regarding the magnetic dipole moments, given in Tables~\ref{tab:hafnium_eismagmom} and \ref{tab:n106isot_eismagmom} for the lowest-energy neutron and proton $K^{\pi}\!=\!6^+$ and $K^{\pi}\!=\!8^-$ configurations, we observe a rather stable behavior of the values calculated for the two-proton configurations. Indeed, along the $Z=72$ isotopic chain, $\mu_{\rm th}^{p}$ ranges for the $K^{\pi}=6^{+}$ states roughly from 5.7~$\mu_{N}$ to 5.9~$\mu_{N}$, and for the $K^{\pi}=8^{-}$ states roughly from 7.3~$\mu_N$ to 7.4~$\mu_{N}$. Similarly, along the $N=106$ isotonic chain, $\mu_{\rm th}^{p}$ ranges for the $K^{\pi}=8^{-}$ states from about 7.3~$\mu_N$ to about 7.6~$\mu_{N}$. This stability results from the fact that, for each set of $K^{\pi}$ quantum numbers, the same proton blocked states are involved in all considered nuclei. A stable behavior is also observed in the values calculated for the $K^{\pi}=8^{-}$ two-neutron configuration along the $N=106$ isotonic chain, again because the same neutron blocked states are involved in the corresponding nuclei.

On the contrary, along the Hf isotopic chain, the calculated magnetic dipole moment of the lowest-energy two-neutron $K^{\pi}\!=\!6^{+}$ and $K^{\pi}\!=\!8^{-}$ states exhibit some jumps which are due to changes in the blocked configurations. A sizable negative value $\mu_{\rm th}^{n} \!\sim\! -1.3$~$\mu_{N}$ is more precisely found for the $K^{\pi}=6^{+}$ two-neutron $(\frac52^+[642],\frac72^+[633])$ configuration in the $^{168-172}$Hf isotopes, whereas an almost vanishing value of $\mu_{\rm th}^{n}$ is obtained for the $K^{\pi}=6^{+}$ two-neutron $(\frac52^-[512],\frac72^-[514])$ configuration from $^{174}$Hf up to $^{178}$Hf, and finally a rather large negative value $\mu_{\rm th}^{n} \sim -2$~$\mu_{N}$ is found for the $K^{\pi}=6^{+}$ two-neutron $(\frac52^-[512],\frac72^-[503])$ configuration in the $^{182-186}$Hf isotopes. Similarly, $\mu_{\rm th}^{n}$ takes a rather small value (in $\mu_N$ unit) between about 0.06 and 0.3 for the $K^{\pi}=8^{-}$ two-neutron $(\frac72^-[514],\frac92^+[624])$ configuration in $^{168-178}$Hf isotopes, whereas it is large and negative around $-2$~$\mu_N$ for the $K^{\pi}=8^{-}$ two-neutron $(\frac72^-[503],\frac92^+[624])$ configuration in the $^{182-186}$Hf isotopes.

The almost vanishing values of $\mu_{\rm th}^{n}$ found for the $K^{\pi}\!= 6^{+}$ two-neutron $(\frac52^-[512],\frac72^-[514])$ configuration in $^{174-178}$Hf, can be ascribed to a cancellation mechanism explained in detail in Ref.~\cite{PRC22_hfbcs_isomers}. On the one hand, the two blocked neutron orbitals have approximately opposite spin projections on the symmetry axis, so that the spin contribution to the intrinsic part $\mu_{\rm intr}$ of $\mu_{\rm th}^{n}$ almost vanishes. This opposite spin content in the two-neutron blocked configuration is visible in the dominant Nilsson quantum numbers $\frac52^-[512]$ (spin up) and $\frac72^-[514])$ (spin down) for the $K^{\pi}=6^+$ states for example. Note also that, for neutrons, the orbital-momentum contribution to $\mu_{\rm intr}$ is zero. On the other hand, the collective contribution $\mu_{\rm coll}$ to $\mu_{\rm th}^{n}$ is positive of the order of a few tenths of $\mu_N$. One thus concludes that the net spin contribution from the intrinsic part of the magnetic dipole moment is slightly negative and has approximately the same absolute value as $\mu_{\rm coll}$. This cancellation mechanism is less pronounced in the $K^{\pi}=8^{-}$ two-neutron states of $^{168-178}$Hf and in the considered $N=106$ isotones (see Table~\ref{tab:n106isot_eismagmom}) with $\mu_{\rm th}^{n}$ between about 0.06 and 0.35 (in $\mu_{N}$ unit).
\\[ -2.2ex]

On the other hand, sizable (or even large) negative values of $\mu_{\rm th}^{n}$ testify to a spin-aligned two-neutron blocked configuration, as can readily be seen in the dominant Nilsson quantum numbers in the  $(\frac52^+[642],\frac72^+[633])$ (spins up) configuration  for the $K^{\pi}=6^+$ isomeric state in the lighter Hf isotopes, in the $(\frac52^-[512],\frac72^-[503])$ (spins down) configuration for the $K^{\pi}=6^+$ state in the heavier Hf isotopes, and in $(\frac72^-[503],\frac92^+[624])$ (spins up) for the $K^{\pi}=8^{-}$ state in the $^{182-186}$Hf isotopes.

Overall we can say that the calculated value for the magnetic dipole moment in a considered two-quasiparticle configuration can strongly constrain the choice of the corresponding neutron or proton blocked states whenever relevant experimental information is available. In this context, the comparison  with $\mu_{\rm exp}$ of $\mu_{\rm th}^{n}$ and $\mu_{\rm th}^{p}$ can help confirming the configuration of an isomeric state. This is e.g.\ the case of the $K^{\pi}=6^+$ isomer in $^{172,174,178}$Hf isotopes which  all have a measured magnetic moment of about 5.6~$\mu_N$, that is very well reproduced by the two-proton configuration $(\frac52^+[402],\frac72^+[404])$. Moreover these isotopes have an excitation energy of about 1.6~MeV, overestimated by about 600~keV in our calculations. Although the lowest-lying two-neutron configurations provide a significantly better agreement with $E^*_{\rm exp}$ in $^{172,174,178}$Hf isotopes, they have a magnetic dipole moment in strong disagreement with the experiment. Therefore the $6^+$ isomeric state of $^{172,174,178}$Hf isotopes is most likely of pure proton two-quasiparticle character. In the same spirit, the comparison of $\mu_{\rm th}^{n}$ and $\mu_{\rm th}^{p}$ with $\mu_{\rm exp}$ for the $8^-$ isomeric state in $^{176}$Yb favors the interpretation of a neutron two-quasiparticle configuration for this state. Indeed, although $\mu_{\rm th}^{n}$ has the wrong sign, it differs much less from $\mu_{\rm exp}$ than $\mu_{\rm th}^{p}$ does. Moreover the calculated excitation energy of the neutron configuration agrees very well with experiment.

The case of the $K^{\pi}=8^-$ isomeric state in $^{178}$Hf is particularly interesting for two reasons. On the one hand, the measured magnetic dipole moment $\mu_{\rm exp} = 3.09$~$\mu_{N}$ considerably differs from the one measured in the neighbouring even Hf isotopes, namely $\mu_{\rm exp} = 7.93$~$\mu_{N}$ in $^{176}$Hf and $\mu_{\rm exp} = 8.7$~$\mu_{N}$ in $^{180}$Hf. On the other hand, $\mu_{\rm th}^{n}$ and $\mu_{\rm th}^{p}$ calculated respectively for the lowest-energy neutron and proton $K^{\pi}=8^-$ configurations in $^{178}$Hf strongly deviate from the measured magnetic dipole moment. At the same time the calculated excitation energies $E^{*n}_{\rm th}$ and $E^{*p}_{\rm th}$ are both in fair agreement with experiment. Since the experimental value $\mu_{\rm exp}$ is kind of {\it sandwiched} between our theoretical estimates $\mu_{\rm th}^{n}$ and $\mu_{\rm th}^{p}$, one could guess that the $8^-$ isomeric state in $^{178}$Hf is a mixture of the two-neutron $\big(7/2^-[514],9/2^+[624]\big)$ configuration and the two-proton $\big(7/2^+[404],9/2^-[514]\big)$ configuration.

We could even attempt to estimate the weight of each configuration by the following simple model. Let us assume that the $8^-$ isomeric state $\ket{\Psi}$ can be written as a superposition $\ket{\Psi} = \alpha \, \ket{\Psi_n} + \beta\,\ket{\Psi_p}$, where $\ket{\Psi_n}$ denotes the SHFBCS solution corresponding to the two-neutron configuration and $\ket{\Psi_p}$ the one corresponding to the two-proton configuration, both  normalized to unity.

The coefficients $\alpha$ and $\beta$ are assumed to be real and to satisfy $\alpha^2+\beta^2=1$. Since the intrinsic magnetic dipole moment operator $\hat{\mu}_z$ commutes with the third component of the isospin operator and because it is a one-body operator, it cannot couple $\ket{\Psi_n}$ and $\ket{\Psi_p}$, so we have
\eq{
  \elmx{\Psi}{\hat{\mu}_z}{\Psi} =
  \alpha^2 \, \elmx{\Psi_n}{\hat{\mu}_z}{\Psi_n}
  + (1-\alpha^2) \, \elmx{\Psi_p}{\hat{\mu}_z}{\Psi_p} \,.
}
One then can estimate the total magnetic dipole moment as
\eq{
  \mu_{\rm th}(\alpha^2) = \alpha^2\mu_{\rm th}^{\rm n}
  + (1-\alpha^2)\mu_{\rm th}^{\rm p}
}
and equating it with $\mu_{\rm exp}$ one can deduce $\alpha^2$ to find
\eq{
  \alpha^2 = \dfrac{\mu_{\rm th}^{\rm p}-\mu_{\rm exp}}
        {\mu_{\rm th}^{\rm n}+\mu_{\rm th}^{\rm p}} \approx 0.56
}
and $\beta^2 = 0.44$. The neutron and proton configurations are therefore estimated to account for respectively 56\% and 44\% to the $8^-$ isomeric state corroborating the assessment in Ref. \cite{KHOO} and references quoted therein.

\section{Summary and concluding remarks}

We explored series of $K^{\pi}=6^{+}$ and $K^{\pi}=8^{-}$ isomeric states in the isotopic and isotonic chains of even-even nuclei around $^{178}$Hf within the Skyrme (SIII) Hartree--Fock--BCS approach. The calculations made for the corresponding neutron and proton two-quasiparticle configurations outline the systematic behavior of the theoretical $6^{+}$ and $8^{-}$ excitation energies in the Hf isotopes and $8^{-}$ energies in the $N=106$ isotones. The comparison with available experimental data provides a test of the SIII Skyrme--Hartree--Fock--BCS approach, as well as, of the considered neutron and proton configurations.

We can outline the following findings with the relevant comments:

(i) In the Hafnium isotopic chain the overall systematic behavior of the calculated $K^{\pi}=6^{+}$ and $K^{\pi}=8^{-}$ two-quasiparticle excitation energies and magnetic dipole moments compared to the available experimental data for the corresponding $K$-isomeric states suggests the dominance of the proton over the neutron configurations. At the same time the model suggests the presence of a mixing between the proton and neutron configurations in the $6^{+}$  excitations in $^{174,178}$Hf and the $8^{-}$ excitations in $^{174,178}$Hf. Also, a possible exception with a dominance of the neutron $6^{+}$ configuration in $^{176}$Hf could be expected as it appears with the lowest energy while the corresponding magnetic moment is experimentally unknown.

(ii) In the $N=106$ isotonic chain the overall systematic behavior of the calculated $K^{\pi}=8^{-}$ two-quasiparticle excitation energies suggests the dominance of the neutron over the proton configurations with an exception in $^{178}$Hf where possible mixing between the two configurations is identified. The latter is supported by the intermediate value of the corresponding experimental magnetic moment while the former (neutron configuration dominance) is supported by the magnetic moment of the $K^{\pi}=8^{-}$ isomer in $^{176}$Yb.

(iii) Regarding the comparison between the theory and experiment, we note that while some discrepancies between the theoretical and experimental energy up to about 300 keV outline the possible limits in the accuracy of the model interpretation (as, e.g., in the $6^+$ isomer of $^{176}$Hf above), the overall systematic behaviour of the $K^{\pi}=6^{+}$ and $K^{\pi}=8^{-}$ isomer energies in the Hf isotopes and $N=106$ isotones is rather well reproduced. At the same time, complementing the model analysis by good reproduction of available magnetic moments data allows for a rather clear identification of the neutron or proton character of the excitation and thus provides a plausible overall pattern for the evolution of the two-quasiparticle $K$-isomers in the Hafnium and trans-lantanide mass region.

(iv) The comparison to results of RHB calculations with DD-ME2 and DD-PC1 functionals in Ref.~\cite{Karakat20} shows a similarity in the structure and energy behaviour of the corresponding two-quasiparticle configurations, as well as in the quality of the model descriptions for the $K^{\pi}=6^{+}$ and $K^{\pi}=8^{-}$ excitations in Hafnium isotopes. More considerable discrepancy between the two approaches is observed for the $K^{\pi}=8^{-}$ excitations in the $N=106$ isotones which calls for further detailed comparison. We note that in \cite{Karakat20} no theoretical values for the magnetic moments in the considered configurations are given, which would be very useful for the better comparison of the two approaches. Also, we remark that a comparison to the HFB approach with Gogny effective interaction as, e.g., applied in \cite{Robledo23} would be very interesting whenever results of corresponding systematic study are available.

(v) The present systematic SHFBCS description of $K$-isomers in the rare-earth and trans-lantanide nuclei, together with the recent model application to heavy and superheavy nuclei \cite{PRC22_hfbcs_isomers}, provide a comprehensive test of both the pairing strengths and the SIII Skyrme parametrization used. Therefore, the overall result opens the way to at least two possible developments: 1) readjustment of the pairing strengths taking into account wide range of isomer energies; 2) adding the $K$-isomer description to a new Skyrme adjustment procedure.

Finally, we conclude that despite some limited imperfections in the relative energies between the s.p.\ states, globally our approach to well deformed rare-earth and trans-lantanide nuclei provides a rather good description of their occurrence around the Fermi energies on two counts: the good reproduction of the general trend of the isomeric energies for the proton configurations under study on one hand, and on the other hand, the occurrence of a mixing of neutron and proton configurations at the right place (at least clearly in the $8^{-}$ case). Moreover, the good reproduction of magnetic dipole moments indicates that their spin structure is rather well described. This is encouraging for further exploration of other well-documented high-$K$ structures in the same mass region of heavy rare-earth nuclei and around.

\begin{acknowledgments}
This work is supported by CNRS and the Bulgarian National Science Fund (BNSF) under Contract No. KP-06-N48/1.
\end{acknowledgments}

\bigskip\bigskip

\end{document}